\documentclass[reprint,twocolumn,aps,pra,amsmath,amssymb]{revtex4-2}
\usepackage{placeins}
\usepackage{graphicx}
\usepackage{dcolumn}
\usepackage{bm}
\usepackage{graphicx} 
\usepackage{placeins}
\usepackage{amsmath}
\usepackage{amssymb}
\usepackage{braket}
\usepackage{xcolor}
\usepackage{comment}

\usepackage[T1]{fontenc}
\usepackage[utf8]{inputenc}
\newcommand{\agp}{\mathcal{A}_\beta}
\newcommand{\expection}[1]{\langle #1 \rangle}

\usepackage{xspace}

\usepackage{listings}
\usepackage{setspace}

\begin{document}

\preprint{APS/123-QED}

\title{Partial Reversibility and Counterdiabatic Driving in Nearly Integrable Systems}

\author{Rohan Banerjee}
\email{rbanerj@bu.edu}
\affiliation{Department of Physics, Boston University, Boston, Massachusetts 02215, USA}

\author{Shahyad Khamnei}
\affiliation{Department of Physics, Boston University, Boston, Massachusetts 02215, USA}

\author{Anatoli Polkovnikov}
\affiliation{Department of Physics, Boston University, Boston, Massachusetts 02215, USA}

\author{Stewart Morawetz}
\email{morawetz@bu.edu}
\affiliation{Department of Physics, Boston University, Boston, Massachusetts 02215, USA}

\begin{abstract}
Adiabatic (or reversible) processes are the key concept unifying our understanding of thermodynamics and dynamical systems. Reversibility in the thermodynamic sense is understood as entropy-preserving processes, such as in the idealized Carnot engine, whereas in integrable dynamical systems it is understood as the conservation of the action variables. Between these two idealized limits, however, things are much less clear. In this work, we study several toy models to assess the extent to which reversible processes are even possible in this regime. We then explore how the dissipative losses resulting from rapidly driving these kinds of systems can be fought by approximate counterdiabatic driving. Finally, we provide numerical evidence that much of the phenomenology should be the same for quantum many-body systems with large degeneracy in the presence of integrability breaking perturbations.
\end{abstract}
\date{\today}
\maketitle

\thispagestyle{plain}


\section{Introduction}
Among the endless variety of non-equilibrium physics, adiabatic (reversible) processes play a special role. In thermodynamics such processes play a pivotal role in defining bounds for efficiency of heat engines or refrigerators, and they have also come to play an important role in quantum computation \cite{nielsenQuantumComputationQuantum2010, albashAdiabaticQuantumComputation2018a}. Due to their reversibility, they do not suffer from the same uncontrolled losses as generic non-equilibrium processes.

The great disadvantage to which adiabatic processes are subject is the requirement of long wait times. This condition will be most familiar to many readers in the quantum context, such as the famous Landau-Zener problem \cite{wittigLandauZenerFormula2005}, where in order to remain adiabatic the driving rate should be smaller than the minimum value set by the energy gap (squared). In classical systems, one can roughly understand the adiabatic condition as requiring that the time it takes for some externally controlled parameter of a system to change by an appreciable amount must be much longer than the time for the system to explore its accessible phase space.

Adiabaticity is straightforward to quantify in two extremes. For those special many-particle systems which are integrable, and can therefore be cast in terms of action-angle variables, each action is a separate adiabatic invariant \cite{Arnold1989}, each of which must be conserved for adiabaticity. This reduces to the well-known $\oint p dx$ in the case of one degree of freedom. On the other hand, for many particle systems which are completely ergodic (a necessary assumption of statistical mechanics), adiabaticity requires conservation of another adiabatic invariant: the volume of phase space enclosed by an energy shell \cite{lochakMultiphaseAveragingClassical1988}, whose logarithm is the more familiar thermodynamic entropy.

In between these two regimes the situation is more complicated. If integrability is broken, but not strongly enough to induce complete ergodicity, there is no universally agreed upon definition of adiabaticity. The conventional definition of adiabaticity is reversibility \cite{landau2013statistical} -- that is, if one executes some slow change of an external parameter of the system, the entropy should not increase. When this change is accompanied by a change in the phase space from regular to mixed (i.e. when breaking integrability), it is not clear \textit{a priori} if such a process is possible without an increase of entropy. The primary focus of our work is to address this question, starting with a classical toy model and then connecting it to quantum many-body systems.

Generically, any finite-time external parameter change will result in diabatic (non-adiabatic) excitations in a system. It is therefore natural to ask the extent to which these diabatic effects can be suppressed. This is the focus of the now-mature field of ``Shortcuts to Adiabaticity'' \cite{Guery-Odelin2019, hatomuraShortcutsAdiabaticityTheoretical2024,duncanTamingQuantumSystems2025, delcampoShortcutsAdiabaticityCounterdiabatic2013}, among which we focus on one in particular: \textit{counterdiabatic driving.} Its application to quantum systems has received extensive attention in recent years, in both an experimental \cite{anShortcutsAdiabaticityCounterdiabatic2016, huExperimentalImplementationGeneralized2018, boyersFloquetengineeredQuantumState2019, meierCounterdiabaticControlTransport2020a,zhouExperimentalRealizationShortcuts2020} and theoretical \cite{selsMinimizingIrreversibleLosses2017,claeysFloquetengineeringCounterdiabaticProtocols2019,cepaiteCounterdiabaticOptimizedLocal2023,petiziolQuantumControlEffective2024,morawetzEfficientPathsLocal2024,schindlerCounterdiabaticDrivingPeriodically2024,takahashiShortcutsAdiabaticityKrylov2024,vanvreumingenGatebasedCounterdiabaticDriving2024,gjonbalajShortcutsAnalogPreparation2025,gangopadhayCounterdiabaticRouteEntanglement2025,duncanExactCounterdiabaticDriving2024,grabaritsFightingExponentiallySmall2025,visuriDigitizedCounterdiabaticQuantum2025,morawetzUniversalCounterdiabaticDriving2025,finzgarCounterdiabaticDrivingPerformance2025,dengisMultimodeNOONstateGeneration2025} context. Although it is less commonly studied in classical settings (see [\onlinecite{Guéry-Odelin_2023}] for a review), it has been employed to find control protocols for perfect adiabatic evolution in special exactly solvable systems \cite{jarzynskiGeneratingShortcutsAdiabaticity2013b, okuyamaClassicalNonlinearIntegrable2016}, approximately in more generic many-body systems \cite{Gjonbalaj_2022}, and more beyond state preparation \cite{vaikuntanathanEscortedFreeEnergy2008,zhongTimeasymmetricFluctuationTheorem2024, cohn-gordonCounterdiabaticHamiltonianMonte2026, campoMoreBangYour2014,tuStochasticHeatEngine2014,villazonSwiftHeatTransfer2019}. Engineering an exact CD protocol \cite{kolodrubetzGeometryNonadiabaticResponse2017}, while possible in a few special cases, is inaccessible in general. This motivated the development of a variational approach for \textit{local} (approximate) CD driving \cite{selsMinimizingIrreversibleLosses2017}, which provides a suitable approximate protocol given a variational ansatz. A dual focus of this work is to study the efficacy of local CD driving in suppressing irreversible losses.


\section{Reversibility \& Integrability Breaking}
Let us begin with a basic question: can perfect reversibility survive when driving breaks integrability? In both integrable and fully ergodic systems, reversibility is achievable if external parameters are varied infinitely slowly. Thus, we classify the reversibility of driving protocols in different systems by their behaviour in the infinitely slow-driving limit. As a testing ground, we define several illustrative toy models whose behaviour in this limit is qualitatively different.

We consider two uncoupled harmonic oscillators with a nonlinear perturbation, one which is integrable and another which is nonintegrable for nonzero perturbation strength $\beta$.

\begin{align}
    H_\mathrm{I} & = \frac{p_x^2 + p_y^2}{2}+\frac{x^2+y^2}{2} + \beta \frac{(x^2 + y^2)^2}{4} \label{eq:HI} \\
    H_\mathrm{NI} & = \frac{p_x^2 + p_y^2}{2}+\frac{x^2+y^2}{2} + \beta \frac{x^2 y^2}{2}\label{eq:HNI}
\end{align}

In each case, the driving protocol consists of ramping $\beta$ from an initial value $\beta_i$ to a final value $\beta_f$. For each protocol, we choose $\beta_i$ and $\beta_f$ such that, in the slow-driving limit, the initial and final average energies coincide across systems, enabling a fair numerical comparison. We define three separate protocols: 


\begin{center}
\begin{tabular}{|m{1.5cm}|m{1.5cm}|m{1cm}|m{1cm}|}
\hline
 Protocol & Model & $\beta_i$ & $\beta_f$ \\ 
 \hline
 I-I & $H_\mathrm{I}$ & 0 & 0.229 \\
 \hline
 I-N & $H_\mathrm{NI}$ & 0 & 1 \\
 \hline
 N-N & $H_\mathrm{NI}$ & 5 & 8.85 \\
 \hline
\end{tabular}
\end{center}

The Hamiltonian in Eq. \eqref{eq:HI} is integrable for all $\beta$, since its radial symmetry guarantees that angular momentum will be a second conserved quantity \cite{Arnold1989}. We define an integrable to integrable (I-I) protocol where this nonlinearity is turned on. The Hamiltonian in Eq. \eqref{eq:HNI} has a phase space structure which depends strongly on the value of $\beta$. For small $\beta$, integrability is only weakly broken and most of the phase space is regular. For large $\beta$, integrability is strongly broken and most of the phase space is chaotic \cite{contopoulosStructureChaosPotential1994,contopoulosDestructionIslandsStability1999,contopoulosNonconvergenceFormalIntegrals2003}. We therefore define an integrable to nonintegrable (I-N) protocol where driving weakly breaks integrability, and a nonintegrable to nonintegrable (N-N) protocol which is within the strongly integrability breaking regime where we expect thermodynamic reversibility to apply. 

To characterize the behaviour of the three systems in the slow-driving limit, we simulate the evolution of an initially microcanonical ensemble with fixed energy $E_0 = 1$ under driving of an external parameter $\beta(t)$. We choose a ramp function, now common in the literature, which is smooth at the beginning and end of the protocol. The key parameter is the total driving time $\tau$, so that $\beta(\tau) = \beta_f$.

\begin{equation}
    \beta(t)=(\beta_f-\beta_i)\sin^2\left[\frac{\pi}{2}\sin^2\left(\frac{\pi t}{2 \tau}\right)\right]+\beta_i
    \label{eq:rampfunction}
\end{equation}

In order to numerically probe reversibility, we need an observable to measure. We take inspiration from quantum mechanics, where if a system begins in a particular eigenstate, non-adiabatic evolution means there will be excitations and hence non-zero fluctuations in final energy. For this reason, we characterize reversibility by the final energy variance of an evolved, initially microcanonical distribution in phase space. This has the advantage of being directly measurable, unlike entropy.

Reversible evolution implies that after any \textit{cyclic} protocol, the state returns back to itself such that the final energy variance must be zero. A cyclic protocol is one in which the external parameter(s), $\beta(t)$ in our case, start and end at the same point. The system is always initialized in a maximum entropy state, where all microscopic configurations consistent with a fixed energy are equally likely. Since entropy cannot decrease, any cyclic protocol with zero energy variance must necessarily have the same entropy as the initial state. Hence, zero energy variance is an indicator of reversibility. In this work, we define a cyclic ramp as one where we first ramp $\beta(t)$ from $\beta_i$ to $\beta_f$ in a time $\tau$, then wait for a randomized time to avoid aliasing effects (see Appendix~\ref{appendix:waiting-time}), and then run the initial ramp in reverse. If the energy variance decays with increasing $\tau$, we say the evolution is reversible. This criterion can also be applied to quantum models.

\begin{figure}
    \centering
    \includegraphics[width = 0.99\linewidth]{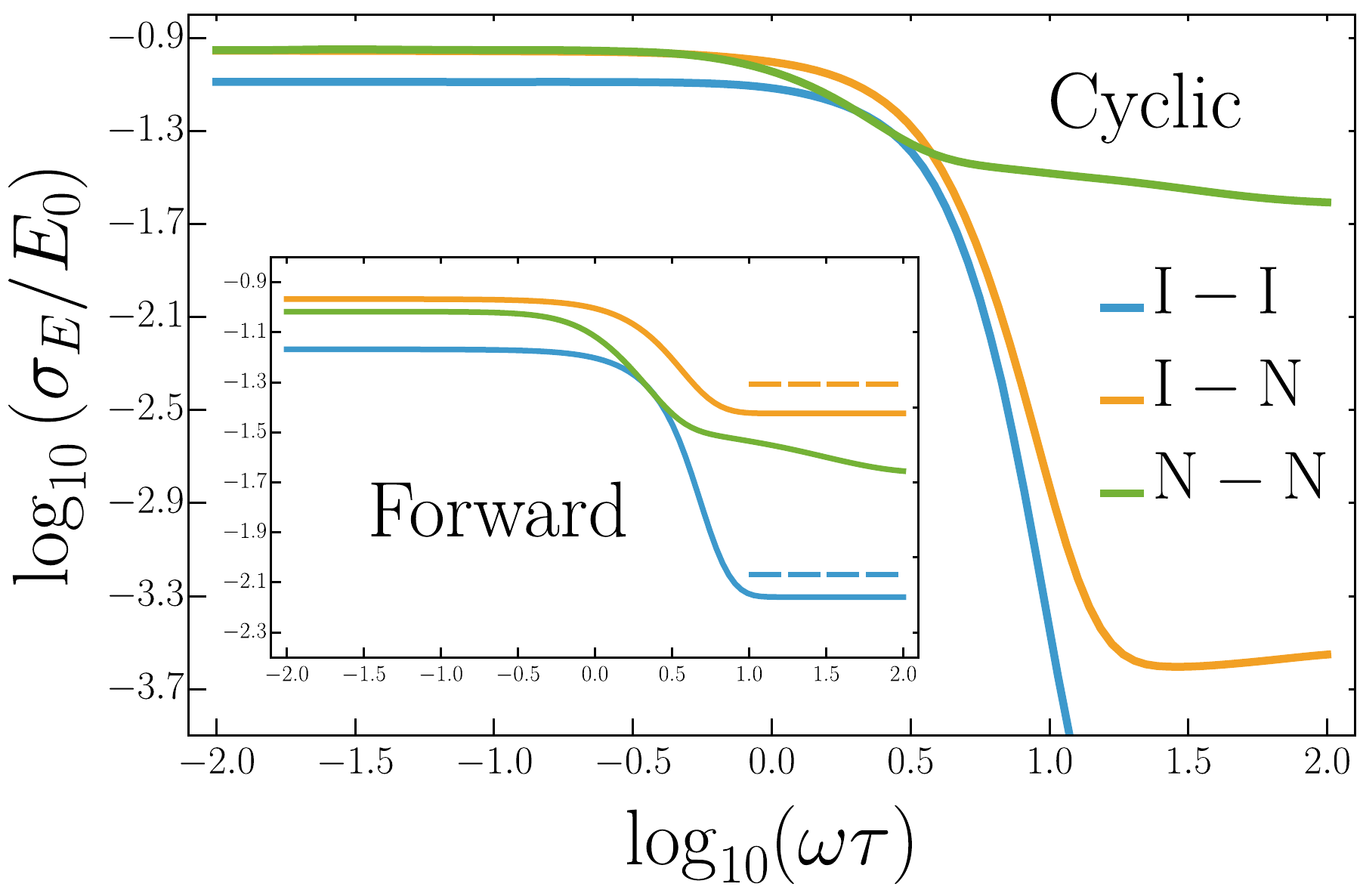}
    \caption{The fluctuations in final energy are shown after evolving an initially microcanonical ensemble with $E_0 = 1$ under various cyclic protocols. Whereas the fluctuations in the extreme cases of totally integrable (I--I) and totally ergodic (N--N) decay in a way consistent with reversibility, the integrability breaking protocol (I--N) plateaus, which we attribute to irreversibility. In the inset, the fluctuations are shown immediately after the initial ramp to $\beta_f$, with estimates given by SW approximation shown as dashed lines.}
    \label{fig:unassisted}
\end{figure}

\begin{figure*}
    \centering
    \includegraphics[width = 1\linewidth]{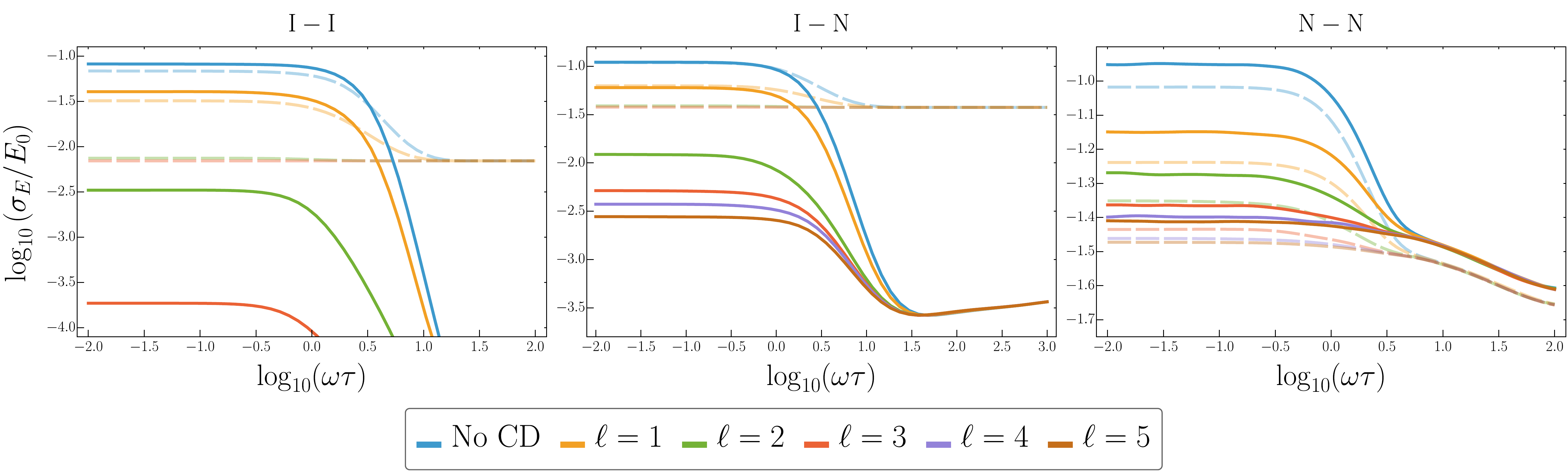}
    \caption{The energy fluctuations induced by the cyclic protocol (solid lines) and after the forward part of ramp is complete at $\beta(t) = \beta_f$ (dashed lines). As the ansatz for local CD driving is increased in complexity (increasing $\ell$), the improvement continues until it reaches a plateau. The fluctuations in the slow driving (large $\tau$) limit provide a lower bound.}
    \label{fig:CD-plots}
\end{figure*}

We show, for each protocol, the energy fluctuations after a cyclic ramp in Fig. \ref{fig:unassisted}. In the two extremes where adiabaticity is well defined -- completely integrable, corresponding to the I--I protocol, and completely ergodic, corresponding to the N-N protocol, we indeed see the fluctuations decaying as the protocol gets longer. The rapid decay in the integrable case, and the very slow decay for the ergodic case is consistent with expectations \cite{landau1976mechanics, ottGoodnessErgodicAdiabatic1979, brownGoodnessErgodicAdiabatic1987}. In the intermediate regime we are interested in, the fluctuations are not suppressed below a plateau, which we attribute to genuine irreversibility. Intuitively, this happens as part of the phase space becomes mixed, and trajectories in that region get scrambled and cannot be uniquely mapped back their starting point.

Quantitatively, one can try to predict fluctuations during the slow protocol by means of the Schrieffer-Wolff transformation \cite{schriefferRelationAndersonKondo1966,bravyiSchriefferWolffTransformation2011}. This is a perturbative method which allows one to compute an effective Hamiltonian $H_{\rm SW}(\beta)$ which approximates $H_0 + \beta V$, but where $H_0$ is an exactly conserved quantity. The fluctuations during the initial ramp of the cyclic protocol -- shown at $\beta(t) = \beta_f$ in the inset of Fig. \ref{fig:unassisted} -- can be computed analytically within this approximation (see Appendix \ref{appendix:SW}), and become asymptotically exact as $\beta\to0$. Since $H_0$ will be exactly conserved to any order in perturbation theory, irreversibility is a non-perturbative effect.


\section{Fighting Dissipation with Local CD Driving}
Next, we study the extent to which these diabatic excitations can be fought by local counterdiabatic (CD) driving. There is a standard derivation of CD driving present in many papers (see e.g. Eq. (62)-(68) of [\onlinecite{kolodrubetzGeometryNonadiabaticResponse2017}]) so we highlight only the main points and refer interested readers to Appendix~\ref{appendix:more-CD} for more detail. The essence of CD driving is to evolve the system according to a modified, \textit{counterdiabatic Hamiltonian}

\begin{equation}
    H_\mathrm{CD}(t) = H(\beta(t)) +\dot{\beta}\agp(\beta(t))
\end{equation}

\noindent where $\agp$ is the \textit{adiabatic gauge potential}. It is responsible for counteracting excitations resulting from external parameters being changed at a finite rate. Evolution under $H_\text{CD}$ is perfectly ``adiabatic'' regardless of how fast $\beta(t)$ is changed.

However, as the exact $\agp$ does not even exist for a generic classically chaotic system \cite{jarzynskiGeometricPhasesAnholonomy1995}, approximations are necessary. Following previous work \cite{claeysFloquetengineeringCounterdiabaticProtocols2019,takahashiShortcutsAdiabaticityKrylov2024,bhattacharjeeLanczosApproachAdiabatic2023}, we use an expansion in Krylov space. Although this was originally devised for quantum systems, it has also been demonstrated \cite{Gjonbalaj_2022} to work well in the classical limit. In this limit, we approximate

\begin{equation}
    \agp^{(\ell)} = \sum_{k=1}^\ell \alpha_k \underbrace{ \{H, \{H, ... \{H}_{2k-1}, \partial_\beta H\}\}\}
    \label{eq:PB_AGP_ansatz}
\end{equation}

\noindent where the $\alpha_k$ are obtained variationally~\cite{selsMinimizingIrreversibleLosses2017}. For reasons of numerical stability, we take a linear combination of these (see Appendix~\ref{appendix:chebyshev-agp}).

The resulting fluctuations after augmenting each protocol with approximate CD driving are shown in Fig. \ref{fig:CD-plots}. The performance is qualitatively similar in each case. When the driving is fast compared to the intrinsic time scale of the system (i.e. $\tau \ll 1 / \omega$), successive orders in the Krylov space expansion can significantly suppress fluctuations, before reaching a plateau. This is consistent with a prediction in recent work \cite{morawetzUniversalCounterdiabaticDriving2025,finzgarCounterdiabaticDrivingPerformance2025}. If the fluctuations in the slow-driving limit are not too small (such as immediately after the forward part of the ramp, see Fig. \ref{fig:CD-plots}) then local CD driving may be sufficient suppress fluctuations as much as the asymptotic $\tau\to\infty$ limit, but it can never beat it.


\begin{figure*}
    \centering
    \includegraphics[width=1\linewidth]{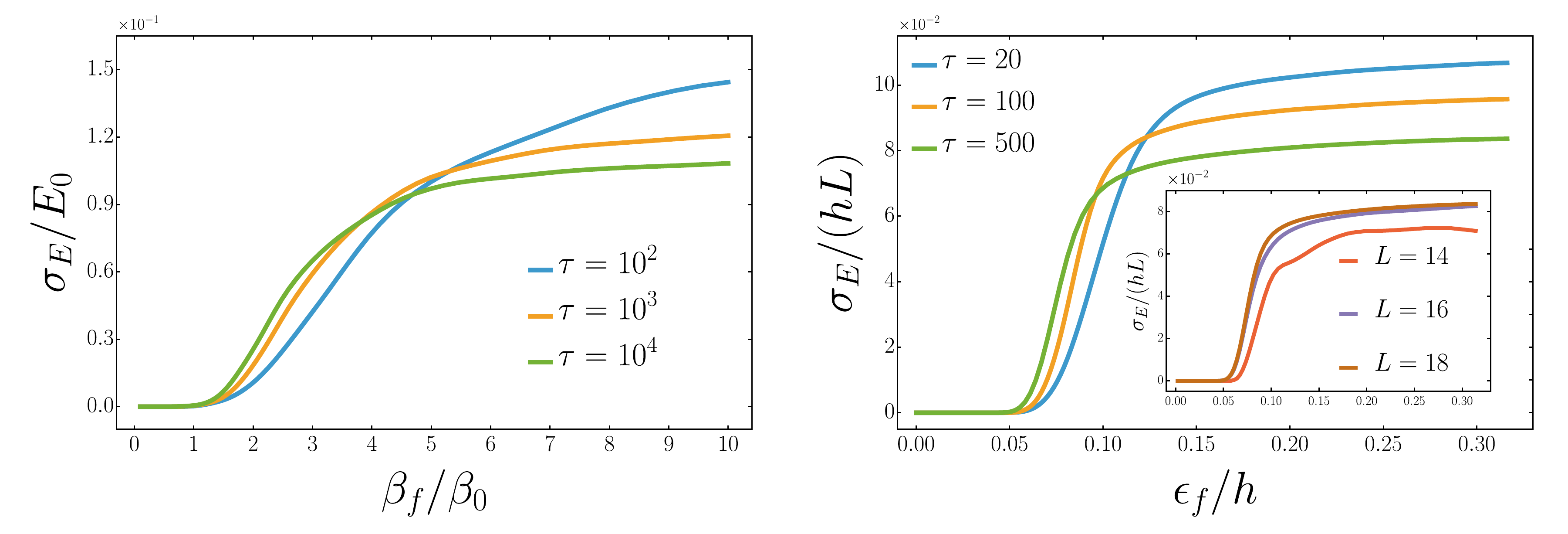}
    \caption{Numerically obtained energy fluctuations after ramping $\beta$ in $H_\mathrm{NI}$ (Eq. \ref{eq:HNI}) and $\epsilon$ in $H_Q$ (Eq. \ref{eq:quantum_model}) to $\beta_f$ and $\epsilon_f$, respectively, and then immediately reversed. In both cases, there are three separate regimes. For $\beta < \beta^\ast$ and $\epsilon < \epsilon^\ast$, energy fluctuations are very small and the protocol is reversible. Beyond this, but within the regime where the SW approximation is valid, we have an ``anti-adiabatic'' effect where slower driving leads to avoided Landau-Zener crossings, and hence greater irreversibility. Eventually, this construction breaks down and slower driving leads to more reversibility.}
    \label{fig:slow-drive}
\end{figure*}

\section{Partial Reversibility in Quantum Systems}
Although so far we have analyzed several classical systems, much of the phenomenology observed should extend to a broad class of quantum or classical many-particle systems, provided they share certain key features. The central ingredient is that the initial, integrable Hamiltonian $H_0$ possesses degenerate symmetry sectors. This property is essential since because one can perform a Schrieffer--Wolff (SW) transformation such that mixing between sectors is eliminated to any finite order in perturbation theory \cite{bravyiSchriefferWolffTransformation2011, WurtzClaeysPolkovnikov2020VariationalSW}.

To numerically demonstrate the phenomenology, we consider the following Hamiltonian: 

\begin{equation} \label{eq:quantum_model}
    \hat{H}_Q = h \sum_\ell \sigma_\ell^z + 4\epsilon \sum_\ell (\sigma_\ell^+ \sigma_{\ell+1}^+ + \sigma_\ell^+ \sigma_{\ell+2}^+ + h.c.)
\end{equation}

\noindent where, like in our harmonic example, $\hat H_0 = h \sum_\ell \sigma_\ell^z$ has a spectrum which with degenerate $U(1)$ symmetry blocks. In Ref.~\cite{abdelshafyOnsetQuantumChaos2025}, it was shown that random matrix level statistics appear in this model for infinitesimally small generic perturbations $\epsilon \hat V$. This was explained as resulting from the fact that the SW Hamiltonian within each block is nonintegrable, and this picture is valid until the split levels in different blocks start to cross.

We will argue that the physics behind partial reversibility is much the same, where non-perturbative effects beyond the SW approximation are responsible for any irreversibility during the I--N protocol. As with the oscillator, we assume that the system begins in a microcanonical ensemble, where all states with a fixed magnetization $M=\sum_\ell \sigma_\ell^{z} = 0$, and hence with a fixed unperturbed energy $E=0$, are equally probable. Within the SW approximation, as the strength of the perturbation $\epsilon$ is increased from $0$, the initially degenerate levels begin to split. There will be some value $\epsilon^\ast$ (see Appendix~\ref{appendix:epsilon-star}) where the split levels in different sectors begin to hybridize, and the SW approximation breaks down. Before this point, i.e. if $\epsilon_f < \epsilon^\ast \ll h$, transitions between the blocks are exponentially suppressed by a finite gap. Hence, for a sufficiently slow cyclic ramp, the absence of transitions between blocks leads to reversibility.

If the integrability breaking perturbation is sufficiently strong that levels hybridize, i.e. $\epsilon^\ast < \epsilon_f \ll h$, global thermalization between blocks sets in. At this point the dynamics becomes irreversible and the cyclic ramp leads to a non-zero energy variance even in the adiabatic limit. An surprising consequence of this analysis is that in this regime we expect that slower ramps should lead to more irreversible behavior, as for faster ramps there is not enough time to transition between different symmetry blocks. This kind of phenomenon has been observed before in the context of Floquet systems \cite{eckardtAvoidedLevelCrossingSpectroscopyDressed2008, weinbergAdiabaticPerturbationTheory2017a}, where slow ramps lead to stronger heating and hence greater irreversibility.

To support these arguments, we perform numerical simulations of the model given in Eq. \eqref{eq:quantum_model}, computed within the zero-momentum block. The dependence of the energy fluctuations after a cyclic protocol on $\epsilon_f$ is shown in Fig. \ref{fig:slow-drive}, for both the classical oscillator (Eq. \ref{eq:HNI}) and quantum spin chain (Eq. \ref{eq:quantum_model}). There are three distinct regimes, common to both systems. Firstly, for $\epsilon < \epsilon^\ast$ energy fluctuations are strongly suppressed, corresponding the reversible regime. As $\epsilon$ is increased beyond $\epsilon^\ast$ but is not too large, irreversibility sets in and we see the predicted ``anti-adiabatic'' region where slower protocols lead to less reversibility. Finally, the perturbative picture breaks down altogether and slower driving leads to more reversibility as expected.


\section{Conclusion}
While adiabaticity is a well-defined notion in systems which are either integrable or ergodic, it is not uniquely defined for the majority of systems, which are neither. Adopting reversibility as a measure of adiabaticity, we study the extent to which slow processes are reversible in this intermediate case. We show that breaking integrability in a system with degeneracies is a fundamentally irreversible process, and provide numerical evidence supporting the claim for both classical and quantum toy models. The underlying reason is that such models can be characterized by approximate emergent symmetries, whose nonperturbative breakdown leads to an associated increase in entropy even in the infinitely slow driving limit. We then test the efficacy of local counterdiabatic driving in the toy model in different regimes, and find that while it can suppress some fluctuations from fast driving, those associated with irreversibility provide a bound that cannot be beaten.

\section*{Acknowledgments}
The authors are grateful to Nachiket Karve and Nik Gjonbalaj for helpful discussions. This work was supported by: NSF (DMR-2412542) and AFOSR (FA9550-21-1-0342). The code and data are available online~\cite{code}. Numerical calculations of classical systems were performed using DifferentialEquations.jl~\cite{Rackauckas2017DifferentialEquations} and QuSpin~\cite{Weinberg2017QuSpin,Weinberg2019QuSpin} was used for quantum systems.

\appendix

\section{SW Approximation for Oscillators}
\label{appendix:SW}

In the main text, we have used the Schrieffer-Wolff transformation to obtain a Hamiltonian for the nonlinear oscillators, unitarily equivalent to the original to any order in perturbation theory, which is completely reversible. We used this to estimate the fluctuations after the forward part of the protocol, shown in the inset of Fig.~\ref{fig:unassisted}. Here, we demonstrate how to perform this calculation, working in the quantum model and then taking the classical limit.

It is more natural to work in terms of creation and annihilation operators $a_x=(x+ip_x)/\sqrt{2}$, $a_y=(y+ip_y)/\sqrt{2}$ (complex wave amplitudes classically), and so we rewrite the harmonic part of the Hamiltonian as
\begin{equation}
    H_0 = \frac{p_x^2 + p_y^2}{2}+\frac{x^2+y^2}{2}=a_x^\ast a_x+a_y^\ast a_y\equiv n_x+n_y.
\end{equation}

The SW transformation is typically performed by writing down the effective Hamiltonian order by order after a unitary transformation, and then inverting a commutator to solve for the required transformation. Here, we use an algebraically simpler method giving the same results. We work in the interaction picture, going to the frame co-moving with $H_0$, which eliminates $H_0$ in the interaction Hamiltonian and maps the remaining operators in $V$ to $a_{x,y}\to a_{x,y} e^{-i t}$, $a^\ast_{x,y}\to a^\ast_{x,y} e^{i t}$. Since there is just a single frequency, one can split separate the perturbation $V$ into Fourier harmonics $V_\ell$ at frequencies $\omega = \ell$. Then, the effective SW Hamiltonian takes the same form of the van Vleck high-frequency expansion (see Refs.~\cite{bukovSchriefferWolffTransformationPeriodically2016, Bukov_2015}) and reads:
\begin{equation}
    H_{\rm SW}=H_0+\beta V_0+\frac{\beta^2}{\hbar}\sum_{\ell>0} \frac{[V_\ell, V_{-\ell}]}{\ell}+\dots,
\end{equation}
where $[A,B]$ is the commutator. Working to leading order in perturbation theory, we keep only the leading non-oscillating term $V_0$, which for the two Hamiltonians we study gives
\begin{align}
    H_\mathrm{I}: \quad  \hat{V}_0 &= \frac{\hbar^2}{8m^2\omega^2} \Bigl((a_x^\dagger)^2 a_y^2 + (a_y^\dagger)^2 a_x^2 \nonumber \\
    &\quad + 3 n_x^2 + 3n_y^2 + 4 n_x n_y + 5 n_x + 5 n_y + 3 \Bigr)\\
    H_\mathrm{NI}: \quad \hat{V}_0 &= \frac{\hbar^2}{8m^2\omega^2} \Bigl((a_x^\dagger)^2 a_y^2 + (a_y^\dagger)^2 a_x^2 \nonumber \\
    &\quad + 4 n_x n_y + 2 n_x + 2 n_y + 1 \Bigr)
\end{align}
Notice that in both cases (and in general) the non-oscillating term $V_0$ commutes with $H_0=n_x+n_y\equiv N$, i.e., $[V_0,H_0]=0$. This emergent symmetry, which we have argued leads to reversibility, exists at all orders of the SW expansion.

If we model the evolution as being adiabatic with respect to changes in $\beta$, then we can compute the energy fluctuations at finite $\beta$ by first computing fluctuations of $V_0$:
\begin{equation}
\begin{split}
    \sigma_E^2 &= \frac{1}{N} \left( \sum_{n_x=0}^{N} \langle n_x, N-n_x \vert \hat{V}_0^2 \vert n_x, N-n_x \rangle \right. \\
    &\quad \left. - \left(\sum_{n_x=0}^{N} \langle n_x, N-n_x \vert \hat{V}_0 \vert n_x, N-n_x \rangle \right)^2 \right).
\end{split}
\end{equation}

After taking the classical limit $N\to \infty,\;\hbar\to 0$ keeping $\hbar N=n$ constant, we find:
\begin{align}
    H_\mathrm{I}:\quad  \sigma_E^2 &= \frac{1}{720} \frac{E_0^4}{m^4 \omega^8} \beta^2 + O(\beta^3), \\
    H_\mathrm{NI}:\quad \sigma_E^2 &= \frac{7}{2880} \frac{E_0^4}{m^4 \omega^8} \beta^2 + O(\beta^3).
\end{align}

\begin{figure}[t]
    \centering
    \includegraphics[width=1\linewidth]{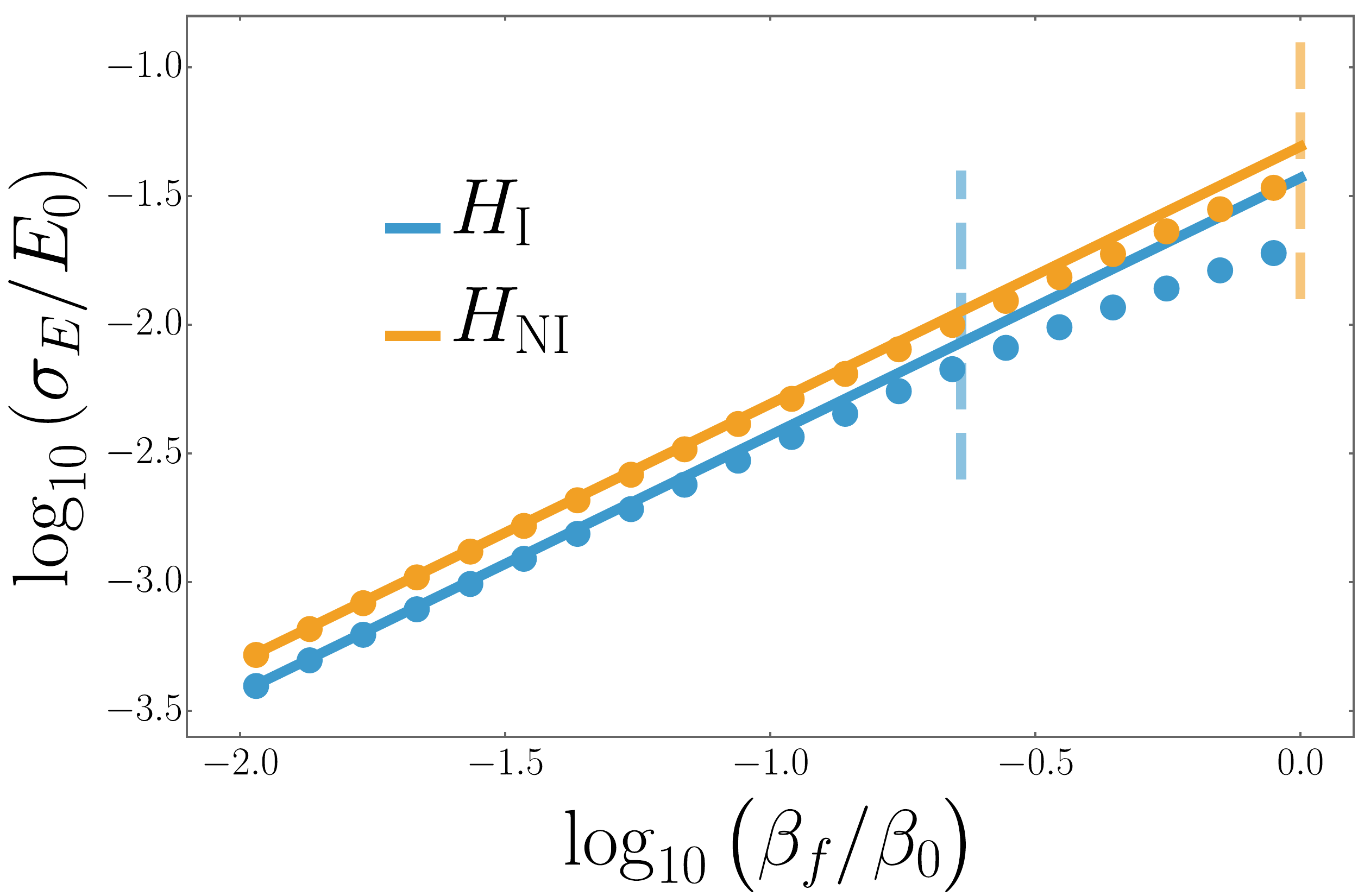}
    \caption{The Schrieffer-Wolff analytical predictions (solid lines), compared with numerical results (dots) for the energy fluctuations from degeneracy-lifting perturbation in the integrable and nonintegrable models. The agreement is excellent at smaller $\beta_f$. The vertical dashed lines indicate $\beta_f$ each protocol.}
    \label{fig:SW-plots}
\end{figure}

We highlight that since these energy variances are obtained within the Schrieffer-Wolff approximation, the physics that they fail to capture is the mixing between different symmetry sectors (different $N$). We compare this analytical prediction with numerical energy fluctuations at the end of the forward ramp for different protocols in Fig.~\ref{fig:SW-plots}. As we expect, it agrees with numerics exceptionally well at small $\beta_f$. We expect that going to higher orders in the expansion would bring them into closer agreement, prior to the ultimate breakdown of the construction when non-perturbative effects become important.

\section{Extracting $\epsilon^\ast$}
\label{appendix:epsilon-star}

As discussed in the main text, when $\hat{H}_Q$ Eq.~\eqref{eq:quantum_model} is driven to increasingly strong perturbations, and $\epsilon_f$ increases beyond $\epsilon^\ast$, the dynamics become irreversible and result in some non-zero $\sigma_E^*$ in the slow-driving limit. For $\hat{H}_Q$, we can roughly calculate $\epsilon^\ast$ as the point at which the typical fluctuations (roughly the standard deviation) of the energy spectrum within a block is equal to the spacing between blocks, i.e., when $\sigma^*_E\sim h$. To the lowest order in the SW approximation:
\begin{equation}
\begin{aligned}
    \hat{H}_\text{SW} = {} & \left(h+128\frac{\epsilon^2}{h}\right)\sum_\ell\sigma^z_\ell \\
    & + 16\frac{\epsilon^2}{h}\sum_\ell\sigma^z_\ell \left(\sum_{j_\ell<k_\ell} \sigma_{j_\ell}^+\sigma_{k_\ell}^- + \sigma_{j_\ell}^-\sigma_{k_\ell}^+\right),
\end{aligned}
\end{equation}
where $j_\ell,k_\ell \in \{\ell-2,\ell-1,\ell+1,\ell+2\}$. Within this approximation, the energy fluctuations within the zero magnetization sector are:
\begin{equation}
    \sigma^2_E = \frac{1}{D}\sum_m\bra{m}\hat{H}_{\rm SW}^2\ket{m} - \left(\frac{1}{D}\sum_m\bra{m}\hat{H}_{\rm SW}\ket{m}\right)^2,
    \label{appendix:variance}
\end{equation}
where $m$ is an eigenstate in the zero magnetization block and $D$ is the number of states in the block. Substituting in $\hat{H}_{\rm SW}$ gives
\begin{equation}
    \sigma_E^2 = 12288 L\frac{\epsilon^4}{h^2}.
    \label{eq:epsilon^*:evar}
\end{equation}

Following the above procedure for estimating $\epsilon^\ast$, if we take it to be the point where $n$ standard deviations of the energies within a block is (half) the inter-block spacing $h$, we obtain:
\begin{equation}
    \epsilon^* = \left(\frac{h}{\sqrt[4]{12288 L}}\right)\frac{1}{\sqrt{n}}.
\end{equation}

For the parameters considered in the numerics shown in Fig.~\ref{fig:slow-drive}: $L = 18$ and $h= 1$, this gives $\epsilon^\ast \approx 0.046 / \sqrt{n}$, which is consistent with the onset of irreversibility seen in Fig.~\ref{fig:slow-drive}. This leads to a very slow finite-size scaling of the point at which irreversibility sets in. A very slow decay to zero is consistent with the numerics that we see in e.g. Fig.~\ref{fig:slow-drive}, although it is difficult to extract the exponent numerically.


\twocolumngrid

\section{Wait time for cyclic protocol}
\label{appendix:waiting-time}

\begin{figure}[b]
    \centering
    \includegraphics[width=1\linewidth]{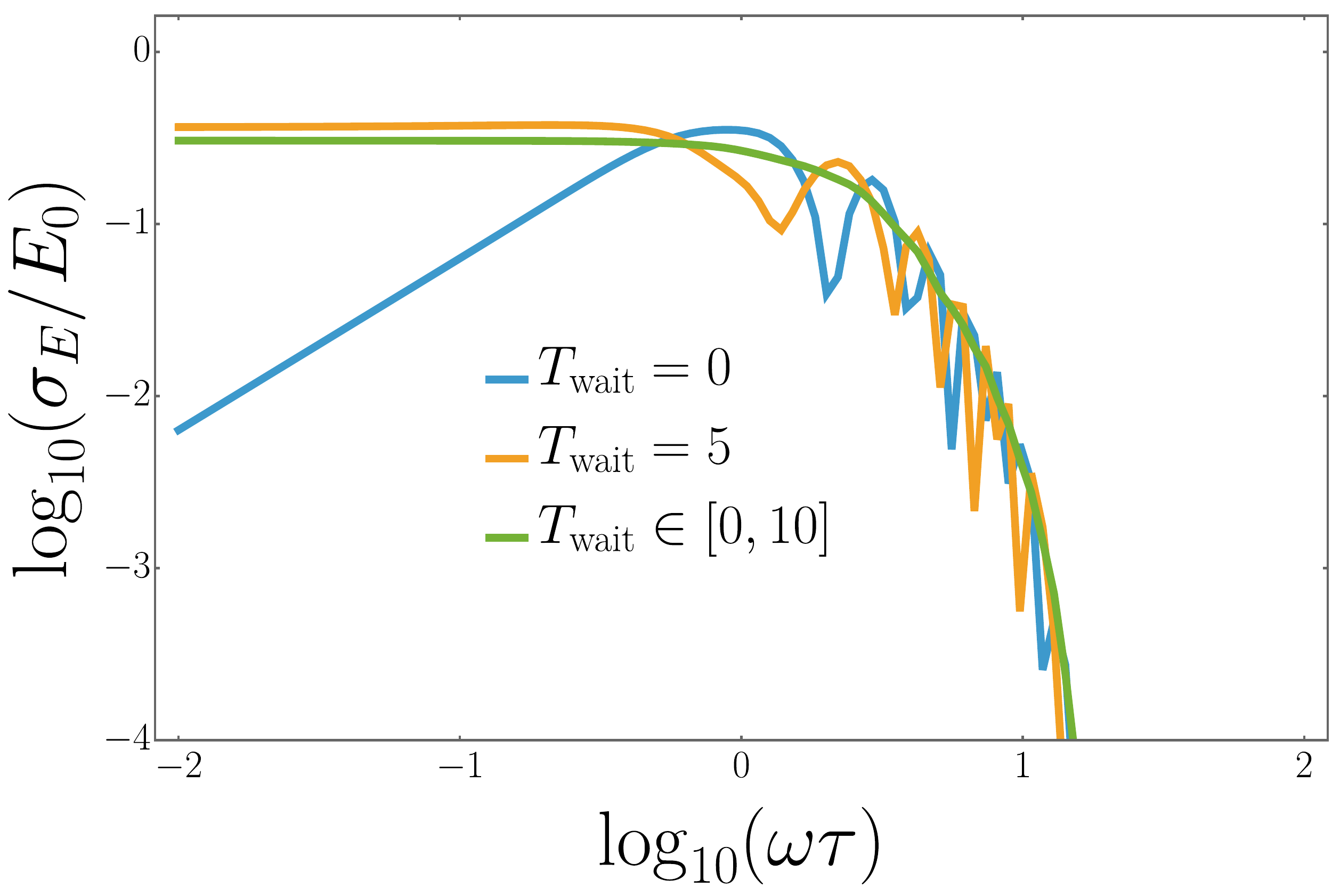}
    \caption{We perform the cyclic I-I protocol with a variety of waiting times. If there is no waiting time (blue line), then we fail to capture the diabatic effects of fast driving. If the waiting time is not random, then we get oscillations corresponding to the fact that the system has some intrinsic period. If the waiting time is randomized, there is a smooth exponential decay of the fluctuations in the slow driving limit, as expected.}
    \label{fig:wait_plot}
\end{figure}

For the slow evolution without any counterdiabatic driving, shown in Fig.~1 of the main text, we have added a randomized waiting time in the middle of the cyclic protocol. Without this, the irreversibility of fast processes is not captured since the protocol consists effectively of a quench to a strong nonlinearity, and then immediately a quench back, without any time for the trajectories to evolve in the newly accessible phase space. A finite but constant waiting time results in aliasing effects due to periodic motion in the system. The effect of including a waiting time for the I--I protocol is shown in Fig.~\ref{fig:wait_plot}, where it results in smooth (exponential) decay in fluctuations in the slow driving limit, as expected for an integrable system.

\section{Detailed derivation of CD driving}
\label{appendix:more-CD}

It is simplest to derive counterdiabatic (CD) driving within the framework of quantum mechanics and then take the appropriate classical limit. This derivation follows Ref.~\cite{kolodrubetzGeometryNonadiabaticResponse2017}. Consider a system whose dynamics are governed by a time-dependent Hamiltonian $H(\beta(t))$, where $\beta(t)$ is a driving parameter that changes from $\beta_i$ at $t=0$ to $\beta_f$ at $t=\tau$. Suppose the system initially occupies an eigenstate $\ket{\psi_n(\beta_i)}$ of $H(\beta_i)$, and we drive it to $H(\beta_f)$ by varying $\beta(t)$. The adiabatic theorem of quantum mechanics~\cite{Sakurai2017} guarantees that in the limit $\tau \to \infty$, the final state approaches the corresponding eigenstate $\ket{\psi_n(\beta_f)}$, i.e., the system undergoes adiabatic evolution. For any finite $\tau$, however, diabatic (non-adiabatic) transitions generally occur. If instead the system evolves under the \textit{counterdiabatic Hamiltonian}, defined as
\begin{equation}
    H_\mathrm{CD}(t) = H + \dot{\beta}\,\agp,
\end{equation}
no diabatic transitions will occur. $\agp$ is known as the adiabatic gauge potential (AGP). Recall that since $\dot{\beta} \propto 1 / \tau$, in the limit of infinitely slow driving, $\dot{\beta} \rightarrow 0$ so we recover adiabatic evolution governed by the original Hamiltonian. One can show that the AGP satisfies the following relation:
\begin{equation}
    \left[\partial_\beta H - \frac{1}{i\hbar}[\agp,H], H\right] = 0.
    \label{eq:cond}
\end{equation}

What is easy to state may not be easy to solve, however, and beyond a select few simple examples, Eq.~\eqref{eq:cond} is not useful in finding the exact $\agp$. Instead, as was shown in Ref.~\cite{selsMinimizingIrreversibleLosses2017}, solving this equation is equivalent to variationally minimizing an action
\begin{equation}
    S_\beta(\agp) = \| G_\beta\|^2 +\mu^2 \|\agp\|^2\,; \quad G_\beta = \partial_\beta H - \frac{1}{i\hbar}[\agp,H],
    \label{eq:action}
\end{equation}
where we have defined the norm
\begin{equation}
    \|O\|^2 = \expection{O^2} - \expection{O}^2,
    \label{eq:norm definition}
\end{equation}
with $\expection{O} = \operatorname{Tr}(\rho O)$, and $\mu$ is some small number added for regularization in case there are many degenerate solutions.
As $\mu\to 0$, minimization of Eq.~\eqref{eq:action} recovers Eq.~\eqref{eq:cond}.

All of this remains perfectly well defined after the quantum to classical mapping of commutators to Poisson brackets, $\frac{1}{i\hbar}[A, B] \rightarrow \{A, B\}$. Then,
\begin{equation}
    G_\beta = \partial_\beta H -\{\agp,H\}.
    \label{eq:classical_G}
\end{equation}

Under this mapping, operators become functions on phase space, and expectation values become
\begin{equation}
    \expection{A} = \int P(\vec{q},\vec{p},\beta) A(\vec{q},\vec{p},\beta) \, d\vec{q}\,d\vec{p}.
\end{equation}
$P(\vec{q}, \vec{p}, \beta)$ is a distribution over conjugate phase space variables $\vec{q}$ and $\vec{p}$. Formally, in classical systems $\agp$ is the generator of adiabatic transformations~\cite{kolodrubetzGeometryNonadiabaticResponse2017, karveAdiabaticGaugePotential2025, kimDefiningClassicalQuantum2025}. Stated less formally, it specifies how to map between stationary (time-independent) distributions in phase space when an external system parameter $\beta$ is changed.

To compute expectation values, we need to know $P(\vec{q},\vec{p},\beta)$. For the initial point $\beta_i$, we use a microcanonical distribution at energy $E_0 = 1$, which is easily sampled from in two dimensions. Ideally, we would like to choose the distribution which is adiabatically connected to this one at each $\beta$. However, since this is not usually accessible, we follow the method used in Ref.~\cite{Gjonbalaj_2022}, and slowly drive the initial distribution from $\beta_i$ to $\beta_f$ without any CD. This allows us to approximate the stationary distribution corresponding to the instantaneous Hamiltonian $H(\beta)$.

\section{Construction of AGP using Chebyshev Polynomials}
\label{appendix:chebyshev-agp}

For the original ansatz given in Eq.~\eqref{eq:PB_AGP_ansatz}, the typical size of the coefficients should decay exponentially with the order in the expansion if each term is to remain of roughly the same size. This leads to numerical instability, so we instead expand in a basis of Chebyshev polynomials, where the variational coefficients will all be of roughly the same size. The Chebyshev polynomials are defined by the recurrence relation: 
\begin{align}
    T_0(x) &= 1, \\
    T_1(x) &= x, \\
    T_{n+1}(x) &= 2xT_{n}(x) - T_{n-1}(x).
\end{align}
To formulate something in terms of operators, we can define a Super-Liouvillian Operator $\mathcal{L} = \{H,\cdot\}$, which corresponds to the multiplication of the phase space function by $\omega$ in frequency space. Recursive application of this defines a set of operators which are orthogonal with respect to a particular measure:
\begin{align}
    Q_0 &= \partial_\beta H, \\
    Q_1 &= \mathcal{L}\{Q_0\}, \\
    Q_{n+1} &= 2\mathcal{L}\{Q_n\} - Q_{n-1}.
\end{align}

Using this modified basis, we can construct the AGP as follows:
\begin{equation}
    \agp^{(\ell)} = \sum^\ell_{k=1}\gamma_k Q_{2k-1},
    \label{eq:Cheby_ansatz}
\end{equation}
where $\ell$ is the order at which we truncate the expansion, and $\gamma_k$ are the variational parameters computed from minimizing the action in Eq.~\eqref{eq:action}. If Eq.~\eqref{eq:Cheby_ansatz} is plugged into Eq.~\eqref{eq:action}, $S_\beta$ is quadratic in the $\gamma_k$. Then, we can rewrite Eq.~\eqref{eq:action} as a quadratic form:
\begin{equation}
    S_\beta(\gamma) = \begin{pmatrix}1 & \gamma\end{pmatrix} \Sigma \begin{pmatrix}1 & \gamma\end{pmatrix}^T; \quad \Sigma = \begin{pmatrix}
        \Sigma_{00} & \Sigma_{0 \gamma}\\
        \Sigma_{\gamma 0}& \Sigma_{\gamma\gamma}
    \end{pmatrix}
    \label{appendix:matrixequation}
\end{equation}
where $\gamma = (\gamma_1 , ... , \gamma_\ell)$. Here, $\Sigma$ is symmetric and $\Sigma_{ij}$ is the sum of the covariance between the $i$-th and $j$-th basis terms for $\agp$ and $G_\beta$ . Solving $\partial_{\gamma_k}S_\beta = 0$ gives 
\begin{equation}
    \gamma = - \Sigma_{\gamma\gamma}^{-1}\Sigma_{\gamma 0}
    \label{appendix:solution}
\end{equation}
For numerical stability, we take the Moore-Penrose pseudoinverse of $\Sigma_{\gamma \gamma}$ when finding the control for the I-I protocol. One can see from the above that the $\alpha_k$ of Eq.~\eqref{eq:PB_AGP_ansatz} are a linear combination of these $\gamma_k$.

\section{Higher Order CD Protocols}
\label{appendix:Higher-Orders}

\begin{figure}[t]
    \centering
    \includegraphics[width=1\linewidth]{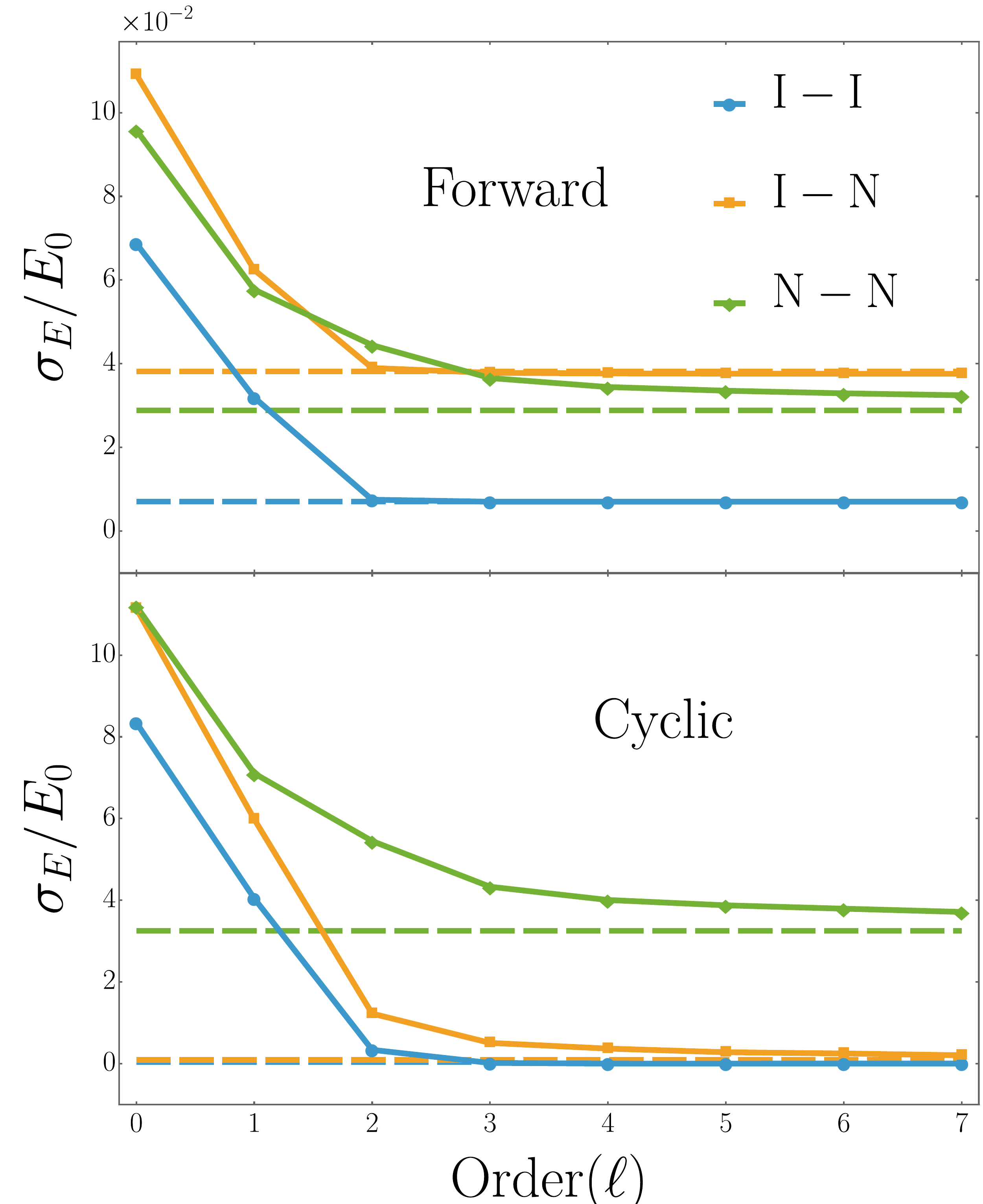}
    \caption{The performance of local CD counteracting fast driving ($\tau = 10^{-4}$) at higher orders in the expansion of Eq.~\eqref{eq:PB_AGP_ansatz}. Dashed lines show the fluctuations from the slow driving limit with $\tau = 10$. The upper panel shows the forward protocol whereas the lower shows the cyclic protocol. In each case, there is a plateau beyond which local CD cannot improve. Depending on the particular protocol, the plateau can still be extremely close to the slow-driving limit result.}
    \label{fig:Higher-Orders}
\end{figure}

In the main text, we show up to five orders in the Krylov space expansion for local CD driving. To show that the performance does indeed plateau, we show the energy variance resulting from up to seven orders in the Krylov space expansion in Fig.~\ref{fig:Higher-Orders}, for both forward and reverse protocols. As mentioned in the main text, this is consistent with previous work suggesting that the improvement from a Krylov space expansion should eventually saturate.

\FloatBarrier
\bibliography{refs}
\end{document}